\documentstyle[11pt,titlepage]{article}
\begin{document}

\newcommand{\be}{\begin{equation}}
\newcommand{\ee}{\end{equation}}

\newcommand{\bea}{\begin{eqnarray}}
\newcommand{\ena}{\end{eqnarray}}
\newcommand{\nn}{\nonumber}

\newcommand{\ta}{{\mbox{\tiny{A}}}}
\newcommand{\tb}{{\mbox{\tiny{B}}}}
\newcommand{\tc}{{\mbox{\tiny{C}}}}
\newcommand{\td}{{\mbox{\tiny{D}}}}

\newcommand{\bz}{{\bar{z}}}
\newcommand{\bc}{{\bar{c}}}
\newcommand{\bh}{{\bar{h}}}

\newcommand{\al}{{\alpha}}
\newcommand{\bt}{{\beta}}
\newcommand{\gm}{{\gamma}}
\newcommand{\dt}{{\delta}}
\newcommand{\eps}{{\epsilon}}
\newcommand{\vep}{{\varepsilon}}
\newcommand{\vp}{\varphi}
\newcommand{\la}{{\lambda}}
\newcommand{\si}{{\sigma}}
\newcommand{\th}{\theta}

\newcommand{\thb}{\bar{\theta}}
\newcommand{\sib}{{\bar{\sigma}}}

\newcommand{\da}{{\dot{\alpha}}}
\newcommand{\db}{{\dot{\beta}}}
\newcommand{\dg}{{\dot{\gamma}}}
\newcommand{\dd}{{\dot{\delta}}}
\newcommand{\dv}{{\dot{\varphi}}}

\newcommand{\La}{\Lambda}
\newcommand{\Lab}{\bar{\Lambda}}

\newcommand{\ca}{{\cal A}}
\newcommand{\cb}{{\cal B}}
\newcommand{\cc}{{\cal C}}
\newcommand{\cd}{{\cal D}}
\newcommand{\ce}{{\cal E}}
\newcommand{\cf}{{\cal F}}
\newcommand{\ch}{{\cal H}}
\newcommand{\cl}{{\cal L}}
\newcommand{\cm}{{\cal M}}
\newcommand{\cn}{{\cal N}}
\newcommand{\cq}{{\cal Q}}
\newcommand{\cv}{{\cal V}}
\newcommand{\cw}{{\cal W}}

\newcommand{\bV}{\bar{V}}
\newcommand{\cwb}{{\bar{\cal{W}}}}

\newcommand{\f}[2]{{\textstyle\frac{#1}{#2}}}

\newcommand{\tr}{{\rm tr}}
\newcommand{\ie}{{\em i.e. }}
\newcommand{\proc}{\, \raisebox{-1ex}{\rule{.2mm}{5mm}}}
\font\fifteen=cmbx10 at 15pt
\font\twelve=cmbx10 at 12pt

\begin{titlepage}

\begin{center}

\renewcommand{\thefootnote}{\fnsymbol{footnote}}

{\twelve Centre de Physique Th\'eorique\footnote{Unit\'e Propre de
Recherche 7061 }, CNRS Luminy, Case 907}

{\twelve F-13288 Marseille -- Cedex 9}

\vspace{3 cm}

{\fifteen  N=2 central charge superspace \\[2mm]
           and a minimal supergravity multiplet}

\vspace{1.4 cm}

{\bf Gernot AKEMANN\footnote{Max-Planck-Institut f\"ur Kernphysik,
                             Saupfercheckweg 1, D-69117 Heidelberg, Germany}, 
     Richard GRIMM, 
     Maximilian HASLER\footnote{D\'epartement de Math\'ematiques,
                                UFR Sciences Exactes et Naturelles,
                                Campus de Fouillole,
                                
\hspace{1.8mm} F-97159 Pointe-\`a-Pitre, Guadeloupe} \\[2mm] 
     and 
     Carl HERRMANN\footnote{allocataire M.E.S.R.}}

\vspace{3 cm}

{\bf Abstract}

\end{center}

We extend the notion of central charge superspace to the case of local
supersymmetry. Gauged central charge transformations are identified as
diffeomorphisms at the same footing as space-time diffeomorphisms and
local supersymmetry transformations. Given the general structure we then
proceed to the description of a particular vector-tensor supergravity
multiplet of $24+24$ components, identified by means of rather radical
constraints.

\vspace{1cm}

\vfill


\bigskip

\bigskip

\noindent December 1998

\noindent CPT-98/P. 3727

\bigskip

\noindent anonymous ftp : ftp.cpt.univ-mrs.fr

\noindent web : www.cpt.univ-mrs.fr

\renewcommand{\thefootnote}{\fnsymbol{footnote}}

\end{titlepage}
\setcounter{footnote}{0}

\indent

{\bf 1.} The problem of gauging central charge transformations in theories with
extended supersymmetry, for the time being mostly $N=2$, has been the subject of
a number\footnote{It is not intended here to give an exhaustive list of
references and we apologize for undue omissions.} of interesting investigations
\cite{Zac78}, \cite{CdWFKST96},
\cite{CdWFKST97}, 
\cite{DT97}, \cite{DT98}, \cite{DIKST98},
\cite{IS98}, \cite{DIKST98}.
In spite of these efforts, the issue of
gauged central charge is far from being appropriately understood and clearly
deserves further study. 
The work presented here is motivated by the superspace
formulation of the $N=2$ vector-tensor multiplet \cite{SSW80a}, \cite{SSW80b}
\cite{dWKLL95}. 
This multiplet exhibits
nontrivial central charge and in \cite{GHH98} it has been shown that central
charge transformations have an interpretation as diffeomorphisms in central
charge superspace \cite{Soh78}, \cite{Gai96}. 
Inspired by this observation we propose here an extension of $N=2$
supergravity to local central charge superspace, employing the usual techniques
of differential geometry in superspace. The frame in this kind of superspace,
\be E^\ca \ = \ dz^\cm E_\cm{}^\ca, \ee
has components $ E^\ca \sim E^a, E^\al_\ta, E_\da^\ta, E^z, E^\bz$. Besides the
components $E^a$, $E^\al_\ta$ and $E_\da^\ta$, which contain the usual
frame (or vierbein) of space-time and the Rarita-Schwinger spinors in their
lowest superfield order, we add components $E^z$ and $E^\bz$
corresponding to the central charge sector. 
The lowest superfield components of $E_m{}^z$ and $E_m{}^\bz$ are then
identified as central charge gauge fields, with local central charge
transformations realised as superspace diffeomorphisms in the central charge
directions. 
These vector fields are candidates to describe the graviphoton of $N=2$
supergravity.

In this general setting one defines the superspace torsion 2-form 
\be T^\ca \ = \ d E^\ca + E^\cb \Phi_\cb{}^\ca, \ee
as the covariant exterior derivative of the frame $E^\ca$. The gauge
connection takes its values in the Lie algebra of the structure group
transformations which, besides the usual Lorentz and $SU(2) \times U(1)$ parts
may now contain nontrivial phase transformations in the central
charge sector with gauge connections $\Phi_z{}^z$ and $\Phi_\bz{}^\bz$,
respectively, as well.

A combined infinitesimal diffeomorphism and structure group transformation 
of parameters $\xi^\cm$ and $\La_\cb{}^\ca$, respectively, changes the frame
as
\be 
\dt E^\ca \ = \ L_\xi E^\ca + E^\cb \La_\cb{}^\ca,
\ee
or, in equivalent covariantized notation :
\be
\dt E^\ca \ = \ \imath_\xi T^\ca + D \xi^\ca 
              + E^\cb \left(\La_\cb{}^\ca - \imath_\xi \Phi_\cb{}^\ca \right),
\ee
Following common usage we define {\em supergravity transformations} as
combinations of diffeomorphisms and field dependent structure group
transformations such that $\La_\cb{}^\ca = \imath_\xi \Phi_\cb{}^\ca$, \ie
\be
\dt E_\cm{}^\ca \ = \ \cd_\cm \, \xi^\ca 
                    + E_\cm{}^\cb \xi^\cc \, T_{\cc \cb}{}^\ca. 
\label{trans}
\ee
Here, the parameters $\xi^\al_\ta, \xi_\da^\ta$ pertain to supersymmetry
transformations whereas $\xi^z, \xi^\bz$ will pa\-ra\-me\-trize gauged central
charge transformations.

\indent

\indent

{\bf 2.} Besides the physical fields of a supergravity multiplet, in the
present case vierbein, Rarita-Schwinger fields and graviphoton, an off-shell
multiplet will also exhibit auxiliary fields.
Different auxiliary field structures of different supergravity multiplets
will reflect themselves in different choices of torsion constraints. 
The general constraints to be described in this section should still allow for
the implementation of different inequivalent multiplets after suitable
supplementary restrictions.
The general constraints consist of two parts.
In a first step we require that the nonvanishing torsion components at
(engineering) dimension zero take the form 
\be 
T {}_\gm^\tc {}_\bt^\tb {\,}^z \ = \ 
     \eps_{\gm \bt} \, \eps^{\tc \tb} c^z, \hspace{1cm}
T {}_\gm^\tc {}^\db_\tb {\,}^a \ = \ 
    -2i \, \dt_\tb^\tc (\si^a \eps)_\gm {}^\db, \hspace{1cm}
T {}^\dg_\tc {}^\db_\tb {\,}^\bz \ = \
      \eps^{\dg \db} \, \eps_{\tc \tb} \, \bc^\bz,
\label{dim0}
\ee
where we allow for $c^z$ and $\bc^\bz$ the possibility to be 
superfields\footnote{Using conventions for chiral $U(1)$ weights such that
$w(E^\al_\ta) = +1, \ w(E_\da^\ta) = -1$, the decompositions employed in
(\ref{dim0}) imply $w(c^z) = -2, \ w(\bc^\bz) = +2$.}, contrary to the flat
case where they were supposed to be constants. These constraints provide
rather mild restrictions and we will not pursue here their
consequences on the structure of the other torsion components.
 
We will instead, in a second step, proceed to implement reality conditions,
motivated by the reality conditions encountered in the
geometrical formulation of the vector-tensor multiplet, but generalized to local
central charge superspace. 
Expressed in terms of torsion components these reality
conditions read
\be c^z T_{z \cb}{\, }^{\underline \ca} \ = \ 
           \bc^\bz T_{\bz \cb}{\, }^{\underline \ca},  
\label{real1}
\ee
where underlined calligraphic indices like $\underline \ca$ run over
ordinary superspace only, \ie $\underline \ca \sim a, {}^\al_\ta, {}_\da^\ta$,
and
\bea
\cd_\cb c^z &=& \bc^\bz T_{\bz \cb}{\, }^z - c^z T_{z \cb}{\, }^z, \\
\label{real2}
\cd_\cb \bc^\bz &=& c^z T_{z \cb}{\, }^\bz - \bc^\bz T_{\bz \cb}{\, }^\bz,
\label{real3}
\ena
in the central charge directions. 
Clearly, these reality constraints are more
stringent than the dimension zero constraints of (\ref{dim0}). 
For instance, taking 
$\cb \sim z$ in (\ref{real1}) yields immediately that the torsion components 
$T_{\bz z}{}^{\underline \ca}$ must vanish. 
A more detailed analysis of the properties of torsions and curvatures subject
to these general constraints will be discussed elsewhere.

Another possible type of further
reduction could be to ask $c^z$ and $\bc^\bz$ to be covariantly constant, \ie 
\be
D c^z \ = \ d c^z + c^z (\Phi_z{}^z - 2 \Omega) \ = \ 0, \hspace{1cm}
D \bc^\bz \ = \ d \bc^\bz + \bc^\bz(\Phi_\bz{}^\bz + 2 \Omega) \ = \ 0.  
\ee
Here $\Omega$ is the $U(1)$ gauge potential, with fieldstrength 
$\Gamma \ = \ d \Omega$, identified in
\be 
\Phi_\bt^\tb{}^\al_\ta \ = \ \dt_\ta^\tb \, \Phi_\bt{}^\al
           + \dt_\bt^\al \, \Phi^\tb{}_\ta 
           + \dt_\ta^\tb \, \dt_\bt^\al \, \Omega,   
\ee
with $\Phi_\bt{}^\al$ and $\Phi^\tb{}_\ta$ the Lorentz and $SU(2)$ gauge
potentials, respectively.
In addition, inspired by the vector-tensor multiplet central
charge transformations one might adopt a parametrization such that
\be 
E^z \ = \ V c^z, \hspace{1cm} E^\bz \ = \ \bar V \bc^\bz,
\ee
for the frame and
$\xi^z = \omega c^z$ and $\xi^\bz  = \bar \omega \bc^\bz$
for the parameters.
In this case $V$ and $\bar V$ inherit the (opposites of the) chiral $U(1)$
weights of $c^z$ and $\bc^\bz$ and one finds
\be
T^z \ = \ F c^z, \hspace{1cm} \mbox{with} \hspace{1cm} F \ = \ d V +2 V \Omega.
\ee
The gauge transformations of $V$ are derived from those of $E^z$ in combination
with the condition $D c^z = 0$. If $\la$ denotes an infinitesimal $U(1)$
transformation, then
\be
\dt \Omega \ = \ - d \la, \hspace{1cm}
\dt V \ = \ d \omega + 2 \omega \Omega + 2 \la V,
\ee
and the transformation law for $F$ reads
\be \dt F \ = \ 2 \omega \Gamma + 2 \la F.\ee
Similar considerations apply to $E^\bz$ and $T^\bz = \bar F \bc^\bz$.
Instead of pursuing a more detailed discussion of these general superspace
structures we shall present, as an example of central charge superspace at work,
the superspace description of a rather restricted supergravity multiplet. 

\indent

\indent

{\bf 3.} The vector-tensor supergravity multiplet is extracted from general
central charge superspace by means of quite {\em radical constraints}. First of
all we restrict the structure group to be the product of Lorentz and $SU(2)$
transformations only (no $U(1)$ and no structure group transformations in the
central charge sector). As to the torsion components we will describe explicitly
a definit  parametrisation in terms of a few superfields, without going into the
details as to what are independent constraints and what are consequences
thereof. To begin with, at dimension zero, the nonvanishing
components are those of (\ref{dim0}), with $c^z$ and $\bc^\bz$ taken to be
constants.  
We then parametrise $E^z = V c^z$ and $E^\bz = \bar V \bc^\bz$. 
Defining $F = dV$ and $\bar F = d \bar V$ we obtain
$T^z = F c^z$ and $T^\bz = \bar F \bc^\bz$. 
The only nonvanishing components, besides $F_{ba}$ and $\bar F_{ba}$, of the
superspace 2-forms $F$ and $\bar F$ are
\be 
F_{\bt \al}^{\tb \ta} \ = \ \eps_{\bt \al} \eps^{\tb \ta},
\hspace{1cm}
\bar F_{\tb \ta}^{\db \da} \ = \ \eps^{\db \da} \eps_{\tb \ta}.
\ee 
At dimension $\f12$ all torsion
components vanish and at dimension one we are left with.
\be
T_{\, \gm \, b \, \da}^{\, \tc \ \, \ta} \ = \ 
        - 2i \eps^{\tc \ta} \, \si^c_{\gm \da} \, \bar F_{cb}, 
\hspace{1cm}
T_{\, \tc \, b \, \ta}^{\, \dg \ \, \al} \ = \ 
        - 2i \eps_{\tc \ta} \, \sib^{c \, \dg \al} \, F_{cb}, 
\label{idf}
\ee
as well as
\be
T_{\, \gm \, b \, \ta}^{\, \tc \ \, \al} \ = \ 
        2 \dt^\tc_\ta \, U^c \, (\si_{cb})_\gm{}^\al, 
\hspace{1cm}
T_{\, \tc \, b \, \da}^{\, \dg \ \, \ta} \ = \ 
        - 2 \dt_\tc^\ta \, U^c \, (\sib_{cb})^\dg{}_\da. 
\ee
This identifies the basic superfields $F_{ba}$, $\bar F_{ba}$ and $U_a$ which
completely describe the components of torsion and curvature, as for instance
the Lorentz curvatures at dimension one,
\be
R_{\dt \gm \, ba}^{\td \tc} \ = \ 
        8 \eps_{\dt \gm} \, \eps^{\td \tc} \, \bar F_{ba}, \hspace{.7cm}
R_{\dt \tc \, ba}^{\td \dg} \ = \ 
        4 \dt^\td_\tc \, (\si^d \eps)_\dt{}^\dg \, U^c \vep_{dcba}, \hspace{.7cm}
R_{\td \tc \, ba}^{\dd \dg} \ = \ 
        8 \eps^{\dd \dg} \, \eps_{\td \tc} \, F_{ba},
\ee 
and the Rarita-Schwinger torsions at dimension $\f32$,
\be
T_{cb \; \ta}^{\ \ \, \al} \ = \ - \cd_\ta^\al \, F_{cb}, \hspace{1cm}
T_{cb \; \da}^{\ \ \, \ta} \ = \ -\cd^\ta_\da \, \bar F_{cb}.
\ee
Moreover we have the chirality conditions
\be 
\cd_\ta^\al \, \bar F_{cb} \ = \ 0, \hspace{1cm}
\cd^\ta_\da \, F_{cb} \ = \ 0,
\ee
and the relations
\be
\cd_\al^\ta U^d \ = \ 
        \f14 \vep^{dcba} \si_{c \, \al \da} \cd^{\da \ta} \bar F_{ba}, 
\hspace{1cm}
\cd^\da_\ta U^d \ = \ 
        \f14 \vep^{dcba} \sib_c^{\da \al} \cd_{\al \ta} F_{ba}.
\ee
As a consequence of the last two equations one obtains
\be \cd^a U_a \ = \ i \vep^{dcba} \bar F_{dc} F_{ba}. \label{divu} \ee
This covariant superspace identity suggests to interprete $U^a$ as the curl
of a two form gauge potential, contributing an antisymmetric tensor to the
multiplet. Its explicit identification requires some more (quite intriguing)
technicalities.

It is however quite obvious to identify all the other component fields in the
superspace geometry presented so far. The vierbein and Rarita-Schwinger fields
are defined as usual,
\ie
\be
E^a \proc \ = \ dx^m e_m{}^a(x), \hspace{1cm}
E^\al_\ta \proc \ = \ \f12 dx^m \psi_{m \, \ta}^{\ \ \, \al}(x), \hspace{1cm}
E_\da^\ta \proc \ = \ \f12 dx^m \bar \psi_{m \, \da}^{\ \ \, \ta}(x).
\label{comp1} \ee
Likewise, the central charge and $SU(2)$ gauge potentials are identified in
\be
\Phi_\tb{}^\ta \proc \ = \ dx^m \, \varphi_{m \, \tb}{}^\ta(x), \hspace{1cm}
V \proc \ = \ dx^m \, v_m(x), \hspace{1cm}
\bar V \proc \ = \ dx^m \, \bar v_m(x).
\label{comp2} \ee
Given these identifications, supersymmetry (or better supergravity
transformations) are obtained using textbook methods. The only missing piece in
this construction is the antisymmetric tensor. 

\indent

\indent

{\bf 4.} The identity (\ref{divu}) has been obtained in the framework of
superspace geometry pertaining to the gravity sector. As it stands it suggests
an interpretation as Bianchi identity for an antisymmetric tensor in the
presence of suitable Chern-Simons forms. In order to obtain a fully
supercovariant description one should embed (\ref{divu}) in the superspace
geometry of a 2-form gauge potential. In what follows we shall derive
(\ref{divu}) from a suitably defined 2-form geometry.

To begin with we consider the superspace 2-form gauge potential $B$ with
invariant fieldstrength $H = dB$ and Bianchi identity $dH = 0$, more
explicitly
\be
E^\ca E^\cb E^\cc E^\cd \left( 
        4 \cd_\cd H_{\cc \cb \ca} 
      + 6 T_{\cd \cc}{}^\cf H_{\cf \cb \ca} \right) \ = \ 0.
\label{bianchi} 
\ee 
Imposing constraints such that all the components of $H_{\cc \cb \ca}$ at
dimension $- \f12$ (all indices spinorial) vanish and at dimension 0 the
nonvanishing components
\be
H_{\gm \tb \, a}^{\tc \db} \ = \ 
          -2i \, \dt^\tc_\tb \, (\si_a \eps)_\gm{}^\db, \hspace{.7cm}
c^z H_{z \tb \ta}^{\ \, \db \da} \ = \ 
          -8 \, \eps^{\db \da} \, \eps_{\tb \ta}, \hspace{.7cm}
\bc^\bz H_{\bz \bt \al}^{\ \, \tb \ta} \ = \
          -8 \, \eps_{\bt \al} \, \eps^{\tb \ta},  
\ee
are all constant, leads, upon repeated use of the Bianchi identities
(\ref{bianchi}), to the identifications
\be U^d \ = \ \frac{i}{24} \vep^{dcba} H_{cba}, \ee
and
\be
\f12 E^\ca E^\cb c^z H_{z \, \cb \ca} \ = \ - 8 \bar F, 
\hspace{1cm}
\f12 E^\ca E^\cb \bc^\bz H_{\bz \, \cb \ca} \ = \ -8F,
\ee
with $\bar F$ and $F$ as identified above in (\ref{idf}).
In fact, the components of $H$ appearing in the last three equations are the
only nonvanishing ones. 
Having made these identifications, the purely vectorial part of (\ref{bianchi}),
\be
\vep^{dcba} \left( 4 \cd_d H_{cba} 
          + 6 T_{dc}{}^z H_{zba} + 6 T_{dc}{}^\bz H_{\bz ba} \right) \ = \ 0,
\ee
reproduces exactly (\ref{divu}). 
Observe that the topological term $\vep^{dcba} \bar F_{dc} F_{ba}$ in this
equation arises automatically. The mechanism presented
here to merge the geometries of the 2-form gauge potential and of supergravity in
superspace is similar to the one used in the geometrical
description of the
$N=1$ new-minimal multiplet. It is in this sense that the multiplet presented
here may be considered as the analogue of new-minimal $N=1$ supergravity. 
The missing
piece in completing the multiplet (\ref{comp1}), (\ref{comp2}), \ie the
antisymmetric tensor gauge field, is now identified in
\be B \proc \ = \ \f12 dx^m dx^n \, b_{nm}(x), \label{comp3} \ee

\indent

\indent

{\bf 5.} Having established the complete superspace geometry relevant for our
multiplet we can now write down the supergravity transformations
\be
\dt_\xi E^\ca \ = \ D \xi^\ca + \imath_\xi T^\ca, \hspace{1cm}
\dt_\xi \Phi_\cb{}^\ca \ = \ \imath_\xi R_\cb{}^\ca, \hspace{1cm}
\dt_\xi B \ = \ \imath_\xi H,
\ee
defined in the usual way as suitable combinations of superspace diffeomorphisms
and field dependent structure group and 1-form gauge transformations. 
Suitable projections of these equations allow to derive supersymmetry as well as
local central charge transformations of the component fields. The supersymmetry
transformations are given as (with usual conventions for summation over spinor
indices)
\bea
\dt e_m{}^a &=& -i \, \psi_m \si^a \bar \zeta - i \, \bar \psi_m \sib^a \zeta, 
\\ [1.2mm]
\dt \psi_{m \, \ta}^{\ \ \, \al} &=& 2 \cd_m \zeta^\al_\ta
     -\frac{i}{6} \zeta^\gm_\ta (\si_{md})_\gm{}^\al 
           \vep^{dcba} H_{cba} \proc
     +4i e_m{}^a \bar \zeta_{\da \ta} \sib^{b \, \da \al} F_{ba} \proc \ , 
\\ [1mm]
\dt \bar \psi_{m \, \da}^{\ \ \, \ta} &=& 2 \cd_m \bar \zeta_\da^\ta
     +\frac{i}{6} \bar \zeta_\dg^\ta (\sib_{md})^\dg{}_\da
            \vep^{dcba} H_{cba} \proc
     +4i e_m{}^a \zeta^{\al \ta} \si^b_{\al \da} \bar F_{ba} \proc \ , 
\\ [1.2mm]
\dt b_{mn} &=& 
       i \, \psi_n \si_m \bar \zeta + i \, \bar \psi_n \sib_m \zeta
            + 4 \bar v_m \, \psi_n \zeta + 4 v_m \, \bar \psi_n \bar \zeta
\ - \ \ m \leftrightarrow n \ , \\ [1.6mm]
\dt v_m &=& \f12 \psi_m \zeta, \hspace{1.5cm}
\dt \bar v_m \ = \ \f12 \bar \psi_m \bar \zeta, \\ [1.6mm]          
\dt \phi_{m \, \tb}{}^\ta &=& 
    \left( \dt^\ta_\td \dt_\tb^\tc - \f12 \dt^\ta_\tb \dt_\td^\tc \right)
    \left\{ 2 i e_m{}^a \left(   
       \zeta_\tc \, \si^b \, \bar T_{ba}{}^\td \proc 
    \, - \,  \bar \zeta_\tc \, \sib^b \, T_{ba}{}^\td \proc \ \right) 
\right. \nn \\ [1.6mm]
       && +4 \zeta_\tc \, \si^{ba} \, \psi_m{}^\td \bar F_{ba} \proc 
       - 4 \bar \zeta_\tc \, \sib^{ba} \, \bar \psi_m{}^\td F_{ba} \proc
\nn \\[1.6mm] 
&& \left. + \f16 ( \zeta_\tc \, \si_d \, \bar \psi_m{}^\td
                 - \bar \zeta_\tc \, \sib_d \, \psi_m{}^\td )
               \, \varepsilon^{dcba} H_{cba} \proc  \ \right\}.
\ena
The supercovariant fieldstrengths appearing in the gravitino transformation law
are identified as usual in the lowest components of the corresponding superspace
tensors, their explicit form being
\be
\bar F_{ba} \proc \ = \ e_b{}^n e_a{}^m 
    \left( \partial_n \bar v_m - \partial_m \bar v_n 
+ \f14 \bar \psi_{n \, \da}^{\ \ \, \ta} 
       \bar \psi_{m \, \ta}^{\ \ \, \da} \right),
\ee
\be
F_{ba} \proc \ = \ e_b{}^n e_a{}^m \left( \partial_n v_m - \partial_m v_n 
+ \f14 \psi_{n \, \ta}^{\ \ \, \al} \psi_{m \, \al}^{\ \ \, \ta} \right),
\ee
and
\be
\vep^{dcba} H_{cba} \proc \ = \ 3 \vep^{dmlk} \left( 
      \partial_m b_{lk} 
       + 16 v_m \partial_l \bar v_k + 16 \bar v_m \partial_l v_k
      + i \psi_m \si_l \bar \psi_k
\right).
\ee
Note the presence of the mixed Chern-Simons form in the fieldstrength of the
antisymmetric tensor. Finally, the covariant fieldstrength of the
Rarita-Schwinger field is given as
\be
T_{cb}{}^\al_\ta \proc \ = \ \f12 e_c{}^n e_b{}^m 
\partial_n \psi_m \, {}^\al_\ta 
-i e_c{}^n \, \bar \psi_{n \, \da \ta} \sib^f{}^{\da \al} F_{fb} \proc
+ e_c{}^n \psi_n \, {}^\bt_\ta (\si_{b e})_\bt{}^\al U^e \proc  
\ \ \ - \ \ \ c \leftrightarrow b ,
\ee
and similar for the complex conjugate fieldstrength.

Although we are working in central charge superspace, nontrivial central charge
transformations are restricted to the sector of $V$, $\bar V$ and $B$. 
Parametrizing
$\xi^z = \omega c^z$ and $\xi^\bz  = \bar \omega \bc^\bz$ one finds
\bea
\dt_{c.c.} V &=& d \omega, \\
\dt_{c.c.} \bar V &=& d \bar \omega, \\
\dt_{c.c.} B &=& -8 \, \bar \omega F -8 \, \bar F \omega.
\ena
The
other components of the multiplet are inert under central charge transformations
due to the drastical torsion constraints we have imposed. 

\indent

\indent

{\bf 6.} One might introduce the combinations $V^\pm = V \pm \bar V$, as well
as  $F^\pm = F \pm \bar F$ and interprete $V^+$ as the graviphoton. Dynamical
equations, \ie Einstein and Rarita-Schwinger equations as well as Maxwell's
second set of equations for the graviphoton are then obtained from imposing
suitable additional constraints. It is not difficult to convince oneself that
that this is achieved through the superfield equations
\be
U_a \ = \ 0, \hspace{1.8cm} 
F^\pm_{ba} \ = \ \frac{i}{2} \, F^{\mp \, dc} \, \vep_{dcba}.
\ee
This trivializes the antisymmetric tensor in the sense that it becomes a pure
gauge. Consistency of this mechanism with the presence of the mixed Chern-Simons
forms in its fieldstrength is ensured by the selfduality conditions. The
$SU(2)$ gauge field becomes pure gauge as well, inducing a covariant gauge with
respect to $SU(2)$ for the gravitini fields.

\indent

\indent

{\bf 7.} We have  presented the general structure of $N=2$ central charge
superspace and emphasized a soldering mechanism involving superspace geometries
relevant for supergravity on the one hand and 2-form superspace geometry,
suitable for the vector-tensor multiplet, on the other hand. Gauged central
charge transformations have an interpretation as a subset of superspace
diffeomorphisms. Correspondingly, the central charge gauge fields appear in the
local frame of superspace. Moreover, this geometric construction implies the
presence of Chern-Simons forms pertaining to the central charge gauge fields
in the fieldstrength of the antisymmetric tensor.

The minimal vector-tensor multiplet we have described is
a particular special case of in this geometrical setting.
In some sense it may be viewed as an analogue of the new-minimal
multiplet in $N=1$ superspace geometry. Contrary to the $N=1$ case,
the emergence of selfduality conditions in the central charge gauge sector
seems to prevent a lagrangean formulation in the present case
(see also \cite{BT98}).
On the other hand, this same property may give rise to relations
with integrable hierarchies, an aspect which deserves further study.

The general geometric setting outlined in the beginning of this paper
("natural constraints" in central charge superspace)
should allow, upon suitable reductions via different types of torsion
constraints, to identify other multiplets, similar to what has been done
in $N=1$ supergravity and in $N=2$ superspace without central
charges. Investigations on these topics are under way.  

{\bf Acknowledgements.} The major part of this work has been done while
G. Akemann visited CPT as a Marie Curie Fellow of the European Commission TMR
Programme (contract nr. ERBFMBICT960997) and M. Hasler benefitted from a
Plassmann Foundation (DFG) Fellowship. We would like to thank C. Klimcik for a
discussion on mixed Chern-Simons forms and selfduality and A. M. Kiss for
checking part of the calculations

\end{document}